# Log Query Interface

Sandeep Singh Sandha[1]     Xin Xu[1]     Yue Xin[1]     Zhehan Li[1]

[1]University of California, Los Angeles

{sandha,xinxu129,yuexin,lizhehan}@cs.ucla.edu

## ABSTRACT

Log Query Interface is an interactive web application that allows users to query the very large data logs of MobileInsight easily and efficiently. With this interface, users no longer need to talk to the database through command line queries, nor to install the MobileInsight client locally to fetch data. Users can simply select/type the query message through our web based system which queries the database very efficiently and responds back to user. While testing on 6GB of datasets our system takes less than 1 seconds to respond back, the similar queries on traditional MySql database takes more than 60 seconds. The system gives user the capability to execute all the queries using sql query language. User can perform complex join operations on very large tables. The query response time is hugely improved by the server side Spark clusters, which stores the big datasets in a distributed system and execute the query in parallel on multiple machines.

## 1. INTRODUCTION

MobileInsight[1] log dataset is useful for researchers from community to do network status analysis, network diagnosis and so on. Currently researchers can only get the log data by raw SQL queries via the mysql database on remote MobileInsight server. However the process of SSH to remote server and then using raw SQL statements brings much inconvenience for researchers and developers. In addition, because of the super large dataset of MobileInsight logs, about 6G, the users need to wait for a long time to get the query results in some cases. And the storage and maintenance of such large dataset is another challenge. These problems can be more tough when the size of the dataset is continuously increasing.

To make all users' access to the data convenient and to deal with the large-data storage and process, we present our work, a web based query application to provide a friendly user interface and accelerate the query speed into milliseconds.

This work contains three independent parts that would be successively involved in when a query comes.

- Web based frontend. When users type in the SQL command (e.g. "SELECT * FROM tMsg"), the JQuery module will handle this message and send the request back to the core service server. This is an essential part for accessing the data conveniently and efficiently. Obviously we don't want the users to interact with the remote server database directly. It is really troublesome to type commands in the terminal to connect the server and input raw SQL statements to do queries on database. These need users to be familiar with SSH and SQL queries. With our friendly front-end, the users can just enter what they want in the website and will get the results shown in good formats.

- Core service server. We use java for our core service server development. When the query request from web based frontend arrives, the core service server should be able to check the format of the query message, and send valid query messages to MySQL server for further queries. So far, we can make the access to data conveniently, but still not efficiently. Thus we introduce the Spark server part. When the java server receives the query from the front-end, it forwards the query to the Spark server to run it on Spark cluster. More specifically, the Spark master runs a Spark-shell and will continuously accept the queries. Then it assigns the tasks to all of its slave nodes.

- Spark server. Spark server is our project's core part. It improves our project's performance a lot. Firstly We deploy our Spark clusters using standalone – a simple cluster manager included with Spark that makes it easy to set up a cluster. When the query request arrives java server for the first time, the java server creates a process to run an interactive Spark shell against the cluster. This Spark shell will be always open and thus the job environment is held. For all the queries coming later, the java server just runs these sql queries on the existing Spark shell process, using Spark sql. Through this interactive Spark shell, we successfully assign all of the queries to the clusters for executing. Once the query result is generated, it will be returned to java server and the java server will parse and display it on the frontend.

We implemented our log query system based on three modes: MySql, Spark on one machine and Spark on a cluster. And we tested their performance separately by running some queries on tFile and tMsg dataset. The tests show that the system running with Spark cluster can make the data querying more than 1000 times faster than running with MySQL and Spark on one machine.

The rest of the paper is organized as follows: Section 2 introduces the background of our Log Query Interface project, section 3 describes the system design of Log Query Interface, section 4 presents our implementation details, section 5 depicts the user flow of our project, section 6 shows evaluation results of systems running with MySql and Spark separately, section 7 reviews related works on the use cases and researches using Apache Spark, section 8 concludes the paper and finally section 9 addresses future work.

## 2. BACKGROUND

The year of 2015 has witnessed the rapid growth of global mobile data traffic. More than half a billion (563 million) mobile devices and connections were added and the data traffic is predicted to reach 30.6 exabytes per month by 2020[3]. The cellular network, which serves as one of the fundamental parts supporting such huge data volume, therefore needs to be researched in-depth.

The cellular network is a very critical infrastructure while it is almost a "black box" to both the applications and users. Typically, mobile applications request for the data transmission through the cellular interface via the socket API without knowing the runtime operations on the bottom layer protocols. And the lack of access into those cellular network protocols makes it difficult for both application developers to design the proper reaction to different network situations and researchers to refine the protocol itself. To break this barrier, UCLA Wireless Networking Group (WiNG) and OSU Mobile System, Security and Networking (MSSN) lab developed a software called MobileInsight.

MobileInsight[1] is a cross-platform package for mobile network monitoring and analysis on end device. It is developed by UCLA WiNG and OSU MSSN lab. Typical usage of MobileInsight consists of declaring a monitor to track the network status and calling analyzer for online/offline analysis. And the data collected from bottom layer of protocol stack by MobileInsight can be shared to InsightShare Plan[2], which aims to build an open and large-scale dataset of cellular networks for the community and by the community.

Currently users from the community can only visit the website of the InsightShare Plan and download the compressed dataset (in .mi2log format). Then use tools embedded in MobileInsight to open and read it, even the developers of MobileInsight have to use raw SQL queries via SSH client to get those data from the MobileInsight mysql database, which is very inconvenient and time-consuming. To address this problem, we develop a user friendly web based query application which makes it easier for even the users from community to get the query results directly from the MobileInsight database.

In addition, now querying the database is really time consuming due to the large data size of MobileInsight logs. For example, simply a SQL command like "SELECT COUNT(*) FROM tMsg" will take the user around 50min to get the result, which is absolutely not acceptable. Actually this database contains more than 245 GB of the records, which means massive data storage, scalability and query efficiency are all big challenges for us. Our team aims to use Spark to deal with these problems. Apache Spark is a very fast and general engine for large-scale data processing. This technology provides an interface for programming entire clusters with implicit data parallelism and fault-tolerance. Spark powers a stack of libraries including SQL and Data Frames which will be used in our project to realize the fast query of the database.

Other than the query accelerating techniques, web-based frontend is also a critical portion of accessing the data. Surely we don't want the users to use a command line interface to interact directly with our database, it's neither intuitive nor extensible, and more importantly, it is not safe under the SQL injection attacks. By using the HTML5 and JavaScript, which have been proved to be mature and applicable technology through the whole Internet, we can not only reduce the cost of being familiar with the database structure, but also build the protection of the backend server.

This query application can help researchers to understand how the cellular network protocol works and possibly find the loophole of current system by presenting collected data in the relation they want. For example, the mobile device may be stuck in 3G network after a handoff while 4G is available. In reality, 62.1% 4G users being stuck in 3G after the call for 39.6 seconds in average[5]. Another loophole in 3G/4G cellular network design would be that in some cases, the mobile user will experience the out-of-service state right after being attached to 4G network due to failure in low-layer (i.e. RRC), which users would never know the reason if the runtime information can not be collected and analyzed in a large scale. Those kinds of loopholes may be found in easier way through the query application.

In all, our team aims to develop a web based query interface for users to do query on the MobileInsight logs conveniently and efficiently. This application's main target users are researchers who need to use the MobileInsight data for analysis.

## 3. DESIGN

Our system (figure 3.1) is composed of a Web front-end interface (written in HTML5, CSS3, JavaScript and jQuery), a Java server, which provides core services to deliver query and query results between the frontend and backend server, a MySQL server and a Spark server. The two backend servers are completely independent of each other. MySQL database is the naive approach to store data, but the performance is not good for large datasets. Therefore, we use Spark Server to provide a distributed data storage and query execution framework to respond to the user in timely manner.

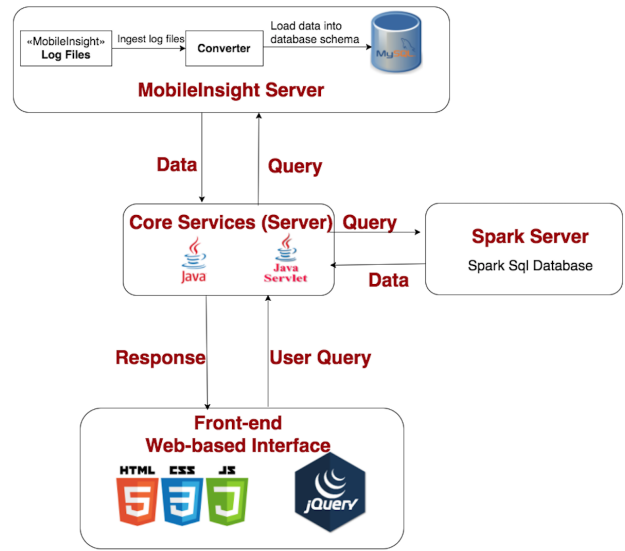

*figure 3.1*

- **Frontend:** we designed the user interface with two main components - query request section and query response section. The query request part is composed of a selection box, which provides a list of query templates and a textbox which allows users to customize their query message. Therefore, users can specify their query messages either by selecting and/or modifying one of the sample queries or customizing it using the textbox. The query results will be displayed in the *Query Result* panel, and the returning data is presented in a table view (figure 3.2).
- **Core services:** We implement a Java server providing core services, which will directly talk to the frontend for requests receiving and results delivering. This middleman acting as a bridge between frontend and data server (i.e. MySQL server or Spark server) can also provide some security functions. It will use identical interface to connect both servers.

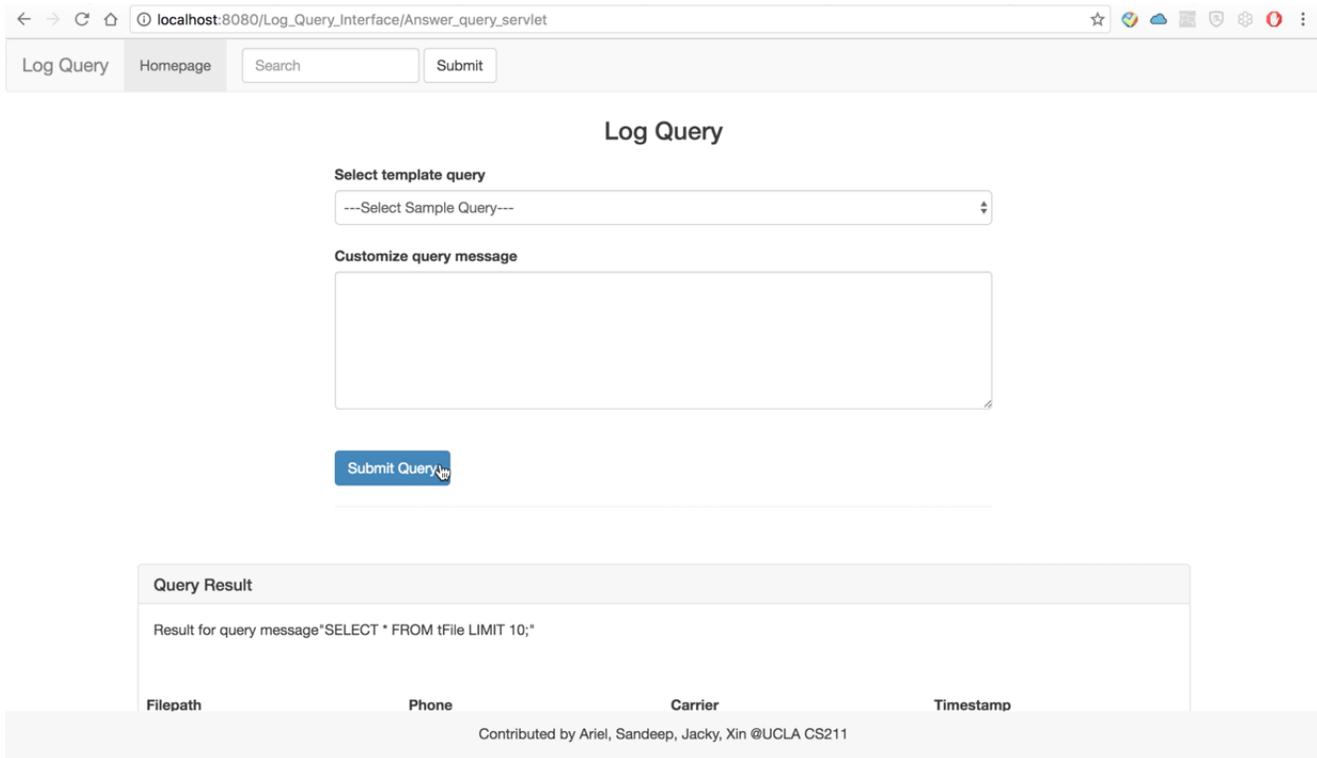

*figure 3.2*

- **MobileInsight server:** This is the part that has been used in the original version of log data query using MySQL server. Now it serves as part of our architecture and the Core service part will send the SQL lines to it and get the results.
- **Spark server:** In particular, our Spark server (figure 3.2) runs on a cluster that involves a cluster manager (i.e. master) and a variable number of work nodes (i.e. slaves). The driver program in Java acts as a coordinator between outside communication and inside cluster, which receives SQL lines and submit the query into the manager. This master will then divide the job into many tasks and deliver them to its slaves, collect the results and send it back to the driver program. Specially, we place the driver program and cluster manager into the same machine (but separated processes) for efficiency even though they play different independent roles in the system. Also, each work node represents a real machine on the same local area network with our Spark server.

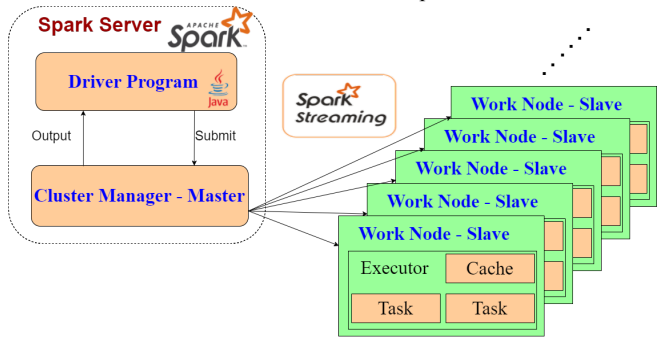

*figure 3.3*

## 4. IMPLEMENTATION

### 4.1 Prerequisite: Eclipse for Java Web development, Java JDK, Apache Tomcat and Spark [4]

**Step 1: Installing Java**

Most machines have Java installed already, use the following command to check the Java version.

```
1  $java -version
```

If Java is already installed on your system, you would see the following response.

```
1  java version "1.7.0_71"
2  Java(TM) SE Runtime Environment (build 1.7.0_71-b13)
3  Java HotSpot(TM) Client VM (build 25.0-b02, mixed mode)
```

Install Java before proceeding to next step in case your system does not have Java installed already.

**Step 2: Installing Scala**

We use Scala language to implement Spark, so you should also verify Scala installation using the following command.

```
1  $scala -version
```

You should see the following response if Scala is already installed on your machine.

```
1  Scala code runner version 2.11.6 -- Copyright 2002-2013, LAMP/EPFL
```

Download the latest version of Scala from http://www.scala-lang.org/download/ if you don't have Scala installed on your system.

Then extracting the Scala tar file using the following command.

```
$ tar xvf scala-2.11.6.tgz
```

Then move the Scala software files to respective directory **(/usr/local/scala)**.

```
$ su –
Password:
# cd /home/Hadoop/Downloads/
# mv scala-2.11.6 /usr/local/scala
# exit
```

Set PATH for Scala.

```
$ export PATH = $PATH:/usr/local/scala/bin
```

Verify the Scala version again using the previous command to make sure that Scala has been properly installed.

**Step 3: Installing Spark**

Downloading the latest version of Spark by visiting https://spark.apache.org/downloads.html .

Extracting the Spark tar and moving the Spark software files to respective directory **(/usr/local/Spark)**

```
$ tar xvf spark-1.3.1-bin-hadoop2.6.tgz
$ su –
Password:

# cd /home/Hadoop/Downloads/
# mv spark-1.3.1-bin-hadoop2.6 /usr/local/spark
# exit
```

Setting up the environment for Spark by adding the following line to **~/.bashrc** file, so then we can locate the Spark software file software using the PATH variable.

```
export PATH = $PATH:/usr/local/spark/bin
```

Use the following command for sourcing the ~/.bashrc file.

```
$ source ~/.bashrc
```

Verify if Spark has been properly installed.

```
$spark-shell
```

**Step 4: Setting Eclipse for Java Web Development [7]**

Then **download Eclipse** by visiting http://www.eclipse.org/downloads/packages/eclipse-ide-java-ee-developers/keplersr2

Import the code into eclipse
1. From the main menu bar, select **File->Import**
2. Select **General->Existing Project into Workspace** and click **Next**
3. Choose either **Select root directory** or **Select archive file** and click the associated **Browse** to locate the directory or file containing the projects.
4. Under **Projects** select the project or projects which you would like to import.
5. Click **Finish** to start the import.

**Install Apache Tomcat server** within Eclipse environment (recommended version: v7). [8]

1. Download Apache Tomcat by visiting http://tomcat.apache.org/.
2. Start the Eclipse WTP workbench.
3. Open **Windows→Preference→Server→Installed Runtimes** to create a Tomcat installed runtime.
4. Click **Add…** to open the **New Server Runtime** dialog, then select your runtime under Apache.
5. Click **Next**, and fill in the Tomcat installation directory.
6. Ensure the selected **JRE** is a full JDK and is of a version that will satisfy Apache Tomcat. If necessary, you can click on **Installed JREs…** to add JDKs to Eclipse.
5. Click **Finish**.

### 4.2 MySQL Server

Create a database:

```
CREATE DATABASE [databasename];
```

Create two tables, tFile and tMsg in the database:

```
USE [databasename];

CREATE TABLE tFile (
    Filepath varchar(255) PRIMARY KEY NOT NULL,
    Phone varchar(20) DEFAULT NULL,
    Carrier varchar(20) DEFAULT NULL,
    `Timestamp` varchar(50) DEFAULT NULL
);

CREATE TABLE tMsg (
    Filepath varchar(255) NOT NULL,
    `Timestamp` varchar(30) NOT NULL,
    MsgType varchar(50) NOT NULL,
    MsgHash varchar(50) NOT NULL,
    MsgPath varchar(255) NOT NULL,
    LineNo varchar(20) NOT NULL,
    PRIMARY KEY (Filepath,`Timestamp`,LineNo)
);
```

Load data(.csv files) to database tables:

```
LOAD DATA INFILE '[filepath]'
INTO TABLE tFile
FIELDS TERMINATED BY ','
LINES TERMINATED BY '\n'
IGNORE 1 ROWS;

LOAD DATA INFILE '[filepath]'
INTO TABLE tMsg
FIELDS TERMINATED BY ','
LINES TERMINATED BY '\n'
IGNORE 1 ROWS;
```

### 4.2 Spark Server

#### 4.2.1 Spark SQL Environment Setup

Apache Spark is a fast and general-purpose cluster computing system, and our Spark server runs as an independent process on the existing cluster. We need to set up this cluster including master and slaves before starting to run the code.

Start a cluster (i.e. a mater and several slaves) manually in the Spark Standalone Mode (http://spark.apache.org/docs/latest/spark-standalone.html). Specifically, execute this command in the master node to start Spark-Mater process:

./sbin/start-master.sh  #Check on the master's web UI: http://localhost:8080 by default.

Similarly, start one or more workers on slave nodes and connect them to the master via:

./sbin/start-slave.sh spark://<master-spark-ip>:7077

After that, check the Spark web UI on your browser: <master-spark-ip>:8080, to make sure you have successfully deployed the whole Spark cluster, which will look as shown in figure 4.1.

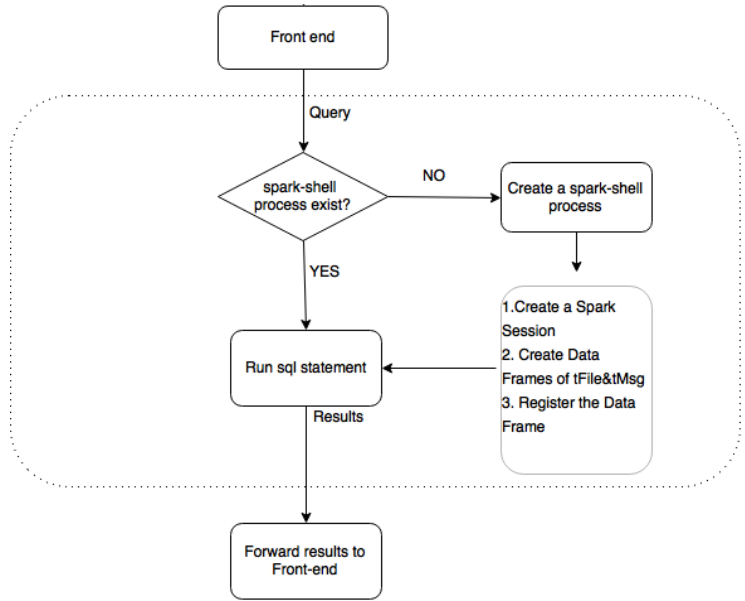

*figure 4.1*

### 4.2.1 Spark Server Logic

The Spark server logic is shown in figure 4.2. When the query request from front-end arrives for the first time, we create a process at runtime to run and start an interactive Spark shell against the cluster. Through this interactive Spark-shell, we can submit our application to the Spark cluster for execution. Here the cluster including Spark master and Spark slave nodes have been deployed before.

We write and run Scala scripts on this shell to do the job:
1. Create a basic Spark Session to enter the all functionality in Spark
2. Create 2 RDDs of tFile objects and tMsg objects from our Mobile-Insight csv file, and convert them to Data Frames.
3. Cache the contents.
4. Register the Data Frames as temporary views.
5. Run the sql statements sent from the front-end using the sql methods provided by Spark.

Then we read the results from Spark cluster and forward them to the front-end.

And for all the queries coming later, the java server uses the existing Spark-shell process and directly runs the new sql statements on it. So we share the same Spark job context/environment within multiple queries to the web server. This is faster than creating a job for each query by keeping data in memory and avoiding reading disks.

### 4.3 Instruction to Run

### 4.3.1 Local Environment Setup

Two changes has to be done in the code for successful running on any environment.

In file:
   *Java Resources/src/Servlet_code/Answer_query_servlet.java*

#### #1: Pointing the storage path of the database files.

Line 95 should points to the tFile and line 105 should points to the tMsg. Correct path for these files need to be substituted. For example for the text file a path of /sparkData/tFile.txt has been shown as example. To use the current code as it is. Put the tFile.txt

*figure 4.2*

and tMsg.txt in the path /SparkData/**fileName* in all the machine used in the cluster. Alternatively HDFS address can be specified here.

#### #2: Pointing the Spark cluster address.

The string command in line 37 points to the Spark installation and initiates a Spark-shell. In this command the Spark installation needed to be updated and the path to the master needs to be added correctly.

### 4.3.2 Run Project

Right click the project folder in Eclipse Workspace→**Run As**→ **Run on Server**

## 5. USER FLOW

With our interactive web frontend interface, users can conveniently do the log query and get quick response from the server because of our big data storage mechanism.

We have two tables stored in our database - tMsg and tFile. Users can query these two tables use normal sql queries, such as SELECT, JOIN, etc.

Our interface has a selection box which contains several template queries, users can pick one at a time as shown in figure 5.1 and figure 5.2.

If the list does not have the query that users are expecting, they can explicitly specify their queries in the "Customize query message" textbox. Users can also modify the selected query message in this textbox as shown in figure 5.3.

User then click "Submit Query" button, and the query result will show up in the panel below. Returning logs are displayed in a table view as shown in figure 5.4.

*figure 5.1*

*figure 5.2*

*figure 5.3*

## 6. PERFORMANCE ANALYSIS

We test our system separately with MySql server, Spark server with one machine, Spark server on cluster with no caching and Spark server on cluster with caching. The size of datasets used by us is 6GB. We used 2 tables to test the system. tMsg and tFile. The structure of these tables is not important. tMsg table is of size 6GB and tFile table is of size 2MB. Table 6.1 gives the summary of the running time of queries.

The queries of table 6.1 are

Query 1: *"Select count(*) from tMsg"*

Query 2: *"Select * from tFile limit 10"*

Query 3: *"Select count(*) from tMsg join tFile on tMsg.Filepath = tFile.Filepath"*

In table 6.1, MySql refers to the system on one machine (4 cores (1.6 GHz) , 4GB RAM) which uses traditional MySql database to answer all the queries. Spark(i) refers to the system which uses Spark on one machine (4 cores (1.6 GHz) , 4GB RAM) without caching. Spark(ii) refers to the Spark on cluster (Consists of 16 slaves: 3 machines each of 2 cores (3.40GHz), 1GB RAM and 13 machines each of 2 cores (2.40GHz), 1GB RAM) without using the in-memory caching capabilities. Spark(iii) is modified version of Spark(ii) where we use in-memory caching.

*figure 5.4*

As seen below MySql and Spark(i) performance is limited due to single machine and using disk to answer the queries. Spark(ii) is order of 10 times faster than MySql and Spark(i) due to using multiple cluster machines to execute the queries. Spark(iii) is the fasted and is order of 100 times faster due to using multiple machines and in-memory processing which is faster than disk. We can see a huge performance improvement with Spark server on cluster.

Query 1 is performing count of the entire tMsg table which is 6GB in size. Thus it testing the reading of entire 6GB of data at runtime to count the number of rows in it.

Query 2, is very short query on tFile table. The size of tFile table is 2MB. Thus this query is very fast and is mostly independent of the number of machine and in-memory processing. The time taken by this query is same in Spark(i), Spark(ii), Spark(iii).

Query 3 is performing a join. Join testing reading of both table and then doing processing on very large datasets of size 6GB. Thus this query test high computation and data access. The speed of this query improves with multiple machine and doing in-memory processing.

MySql is faster than Spark(i) because Spark takes some time to initialize the executor and assign the work.

| Time | Query 1 | Query 2 | Query 3 |
|---|---|---|---|
| *MySql* | 60 Sec | 0.26 Sec | 65 Sec |
| *Spark(i)* | 100 Sec | 0.35 Sec | 102.8 Sec |
| *Spark(ii)* | 5.53 Sec | 0.35 Sec | 5.68 Sec |
| *Spark(iii)* | 0.64 Sec | 0.35 Sec | 0.99 Sec |

*table 6.1*

## 7. RELATED WORK

We apply Spark Apache, which is a fast and general engine for large-scale data processing, to store and process our huge amount of log data. Actually Spark is used at a wide range of organizations to process large datasets.

The teams from Bing, Microsoft need to monitor and analyze user engagement, act upon revenue opportunities in markets around the world. So they use Apache Spark Streaming to collect logs and signals associated every single search query, process and enrich the data in near real-time[9].

AirStream[10] is a realtime stream computation framework built on top of Spark Streaming and Spark SQL. It allows engineers and data scientists at Airbnb to easily leverage Spark Streaming and SQL to get real-time insights and to build real-time feedback loops. There have been a few production use cases such as real-time ingestion pipelines for data warehouse, and computing derived data for online data products.

Baidu's deep learning technology uses Spark to drive deep learning training and prediction using Paddle, the deep learning library developed by Baidu IDL[11]. This enables multiple Baidu's production offline processing to do data ingestion, preprocessing, feature extraction and model training in one Spark cluster. they also address the resource heterogeneity to support multi-tenancy using Spark.

Spark Streaming solves the real-time data processing problem, but to build large scale data pipeline we need to combine it with another tool that addresses data integration challenges. Thus the Apache Kafka project[12] introduced a new tool, Kafka Connect, to make data import/export to and from Kafka easier and bridge the gap between other data systems and stream processing framework.

The teams from Trifacta built a new engine that casts data profiling as an OLAP problem and leverages Spark to quickly generate query results[13]. Its low latency enables 'pay-as-you-go' profiling, empowering users to explore their data iteratively, summarizing columns only as needed and executing focused drill-down queries too expensive to apply broadly. We can see 10x-100x speedups with Spark and faster still in pay-as-you go cases.

Spark Apache is also widely applied in research areas to help solve problems.

VAST Research lab from UCLA presented Blaze[14], an accelerator-aware runtime system that enables rapid warehouse-scale accelerator deployment for the Hadoop/Spark ecosystem. Blaze provides accelerator management for data-intensive scalable computing (DISC) systems in a cluster with heterogeneous accelerator platforms, including GPUs and FPGAs.

The Scientific Computing team at the University of Washington's Institute for Health Metrics and Evaluation is extending the landmark Global Burden of Disease study to forecast what the future of the world's health might look like under a variety of scenarios[15]. In order to do so, they created massively detailed simulations of the world, using Spark to distribute a workload that can produce up to a petabyte of outputs per run.

## 8. CONCLUSION

We presented the design, development and testing of Log Query Interface which is web based system to query very large datasets. Log Query Interface allows users to query the large dataset easily and efficiently. From our final results we saw that the query time can be improved by factor of 100 time using simple machine to form a cluster than using one machine. The main factor driving the development was to process the queries in realtime. We also compared and tested the system using in-memory processing and without using in-memory processing. In our testbed, the in-memory processing improved the performance by 10 times.

## 9. FUTURE WORK

Adding capability to add data in real-time from multiple devices to the database. Figure 9.1 highlights the parts which can increase the capabilities of the current system. The logs can be added to the database in real-time. Thus the query results will be reflected in real-time. Functionalities on Spark server are described below:

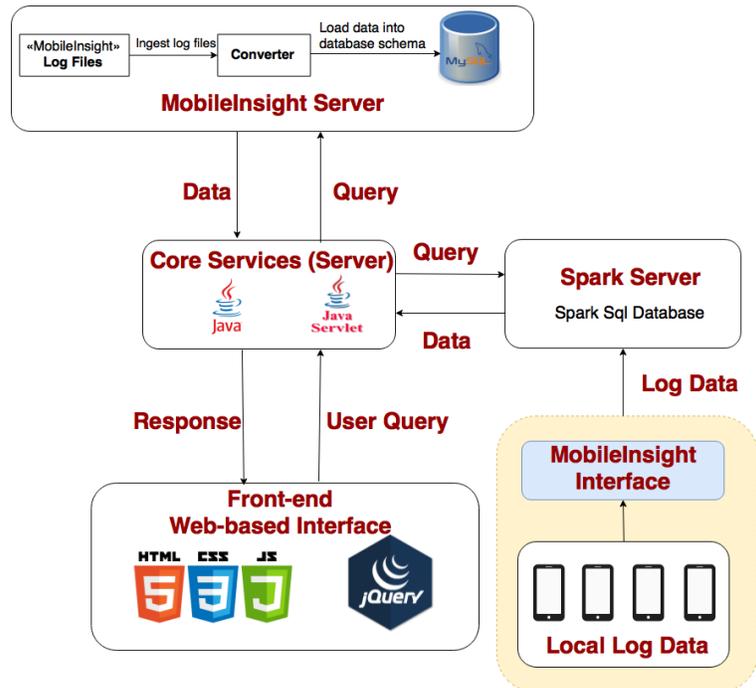

*figure 9.1*

*Refreshing cache of Spark:* As the new data is added at run time, the cache of Spark will need to be refreshed after regular intervals to reflect the latest data. The refreshing rate will decide the data consistency.

*Data curation at runtime:* Since data from multiple sources may not always be correct, data validation will be important component which will require intelligent system design.

*Data Pre-processing:* We might be interested to do the data Pre-processing to make data in the format to be stored, to understand and detect anomalies in the data in realtime. This component will be designed on the basis of particular requirements and semantics of the data.